\documentclass[%
 reprint,longbibliography,preprintnumbers,
nofootinbib,
 amsmath,amssymb,
 aps,
prl,
]{revtex4-1}
\pdfoutput=1
\usepackage[utf8]{inputenc}
\usepackage{flushend}
\usepackage{dcolumn}
\usepackage{bm}
\usepackage{balance}

\usepackage[normalem]{ulem}

\usepackage[colorlinks = true,
            linkcolor = blue,
            urlcolor  = blue,
            citecolor =green,
            anchorcolor = blue]{hyperref}
\usepackage{verbatim}
\usepackage{color,ulem}
\usepackage[english]{babel}

\usepackage[utf8]{inputenc}
\input Starburst.fd
\newcommand*\initfamily{\usefont{U}{Starburst}{xl}{n}}\initfamily

\newcommand{\beq}{\begin{eqnarray}}
\newcommand{\eeq}{\end{eqnarray}}
\usepackage{amsmath}
\usepackage{tikz}
\usetikzlibrary{decorations.pathmorphing}
\usetikzlibrary{shapes.misc}
\tikzset{cross/.style={cross out, draw=black, minimum size=8*(#1-\pgflinewidth), inner sep=0pt, outer sep=0pt},
cross/.default={1pt}}
\usetikzlibrary{patterns,math}
\begin{document}

\title{Molecular-level relation between intra-particle glass transition temperature and stability of colloidal suspensions}

\author{Carmine Anzivino$^{1}$}%
\email{carmine.anzivino@unimi.it,}

\author{Alessio Zaccone$^{1,2}$}%
 \email{alessio.zaccone@unimi.it}
 
 \vspace{1cm}
 
\affiliation{$^{1}$Department of Physics ``A. Pontremoli'', University of Milan, via Celoria 16,
20133 Milan, Italy.}
\address{$^2$ 
I. Physikalisches Institut, University of Goettingen, Goettingen, Germany}

\begin{abstract}
In many colloidal suspensions, the dispersed colloidal particles are amorphous solids resulting from vitrification. A crucial open problem is understanding how colloidal stability is affected by the intra-particle glass transition. 
By dealing with the latter process from a solid-state perspective, we estabilish a proportionality relation between the intra-particle glass transition temperature, $T_{\textrm{g}},$ and the Hamaker constant, $A_\textrm{H},$ of a generic suspension of nanoparticles. It follows that $T_\textrm{g}$ can be used as a convenient parameter (alternative to $A_\textrm{H}$) for controlling the stability of colloidal systems. Within DLVO theory, we show that the novel relationship, connecting $T_\textrm{g}$ to $A_\textrm{H},$ implies the critical coagulation ionic strength (CCIS) to be a monotonically decreasing function of $T_{\textrm{g}}.$ We connect our predictions to recent experimental findings. 
\end{abstract}


\maketitle


Suspensions of colloidal particles dispersed in a liquid solvent represent a paradigm of complex systems where the interplay between a hierarchy of physico-chemical effects across different length and time scales leads to fascinating mesoscopic behaviours, from fractal aggregation \cite{Dimon} to phase separation and phase transitions such as colloidal gelation \cite{Weitz_2008,Gibaud,Eberle,Schreiber,Joep}, vitrification and crystallization \cite{Weitz_AnnualReview}. 

The stability of a colloidal suspension with respect to the aggregation of the dispersed colloidal particles (a phenomenon also known as \textit{coagulation}) is explained by classic Derjaguin-Landau-Verwey-Overbeek (DLVO) theory \citep{Landau,V_O} in terms of a balance between attractive van der Waals (vdW) and repulsive electrostatic double-layer forces, respectively \cite{Sader}. A loss of colloidal stability due to either addition of salt \citep{Landau,V_O} or application of shear flow \cite{shear_1,shear_2} can eventually result in the suspension coagulating rapidly, and into gel formation. The presence of ions in the solution, indeed, can lead to a compression of the double layer as a consequence of which attractive interactions among the colloids dominate the repulsive ones. The threshold of ionic strength above which a colloidal suspension is destabilized due to rapid particle aggregation is known as the \textit{critical coagulation ionic strength} (CCIS) \cite{russel,Borkovec}.

Much research in the last years has concerned the role played on the stability of colloidal suspensions by so-called non-DLVO forces which include, among others, hydration repulsion, hydrophobic interactions and protrusion forces \citep{NINHAM19991}. Comparatively, less attention has been devoted to investigate the role played by the physical properties of the dispersed colloidal particles. 

Apart from a few important exceptions (e.g. colloidal nanocrystals), most widely dispersed colloids, such as silica and polymer nanoparticles, are amorphous solids with  an internal disordered structure. They indeed commonly result from a vitrification process, i. e. a glass transition, with the glass-transition temperature $T_\textrm{g}$ being an important material property that characterizes the degree of mobility within the colloids \citep{mobility}. Understanding how the intra-particle glass transition temperature $T_\textrm{g}$ affects the stability of a colloidal suspension is a fundamental open question. 

In recent experiments, Scott et al. \citep{Priestley} investigated the effect of adding salts on the stability of a series of suspensions of electrostatic-stabilized polymer nanoparticles spanning a wide range of $T_{\textrm{g}}$ values. When adding the hydrophilic KCl salt, these authors found the CCIS to be a decreasing function of $T_{\textrm{g}}.$ A more intricate effect was observed when adding hydrophobic salts. In all cases, an explanation for these findings was suggested in terms of the lifetime of ionic structuration within mobile surface layers, i.e. chiefly in terms of non-DLVO effects \cite{Priestley}.

\begin{figure}
\centering
\includegraphics[width = 0.9 \linewidth]{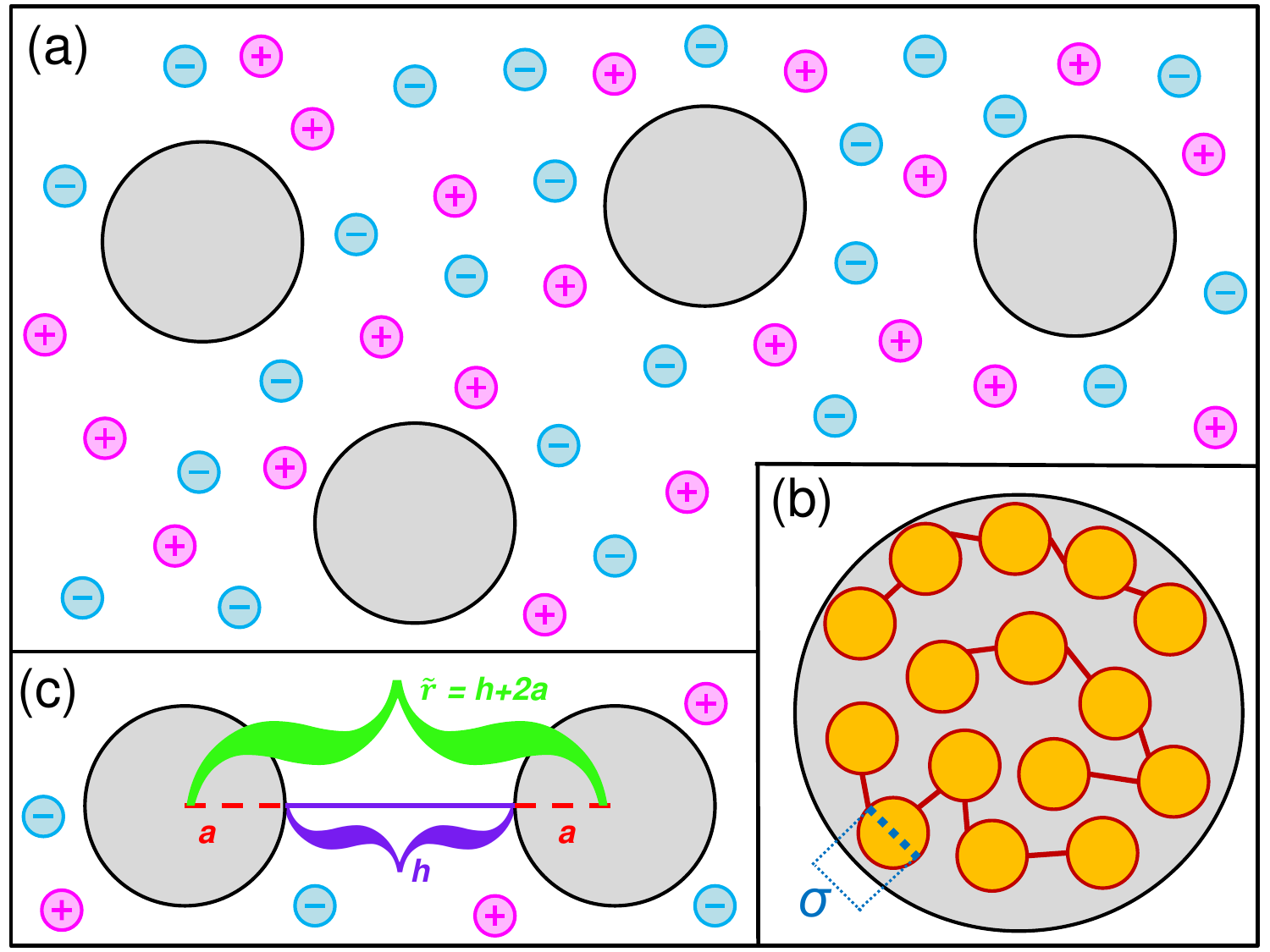}
	\caption{$(a)$ Sketch of the colloidal suspension considered in this Letter. Dispersed colloids have a spherical shape with radius $a$ and, as showed in the zoom $(b),$ consist of many polymer chains vitrified at a temperature $T_\textrm{g}.$ The polymer chains have same length and chemical structure, and the monomer units inside each chain are spheres with radius $\sigma.$ Small magenta and light blue particles indicate positive and negative electolytes, respectively, eventually present in the solution. In $(c)$ a zoom on two particles is considered. While $h$ is the surface distance, $\tilde{r}= h + 2a$ is the center-to-center separation.}
		\label{FIG1}
\end{figure}

In this Letter, we examine this fundamental question from a theoretical point of view. We consider a suspension (see the sketch in Fig. \ref{FIG1}) whose dispersed colloids have a spherical shape and consist of many polymer chains, with same length and chemical structure, vitrified at a temperature $T_{\textrm{g}}.$ The strength of the vdW attraction among the colloids is conveniently taken into account by the so-called Hamaker constant $A_\textrm{H}$ \citep{Hamaker}, which is a material-dependent coefficient providing quantitative information on the London dispersion forces acting among the monomer units of the polymer chains inside each colloidal particle \citep{Hamaker, Israelachvili}.
Following Zaccone and Terentjev \citep{GLASS_THEORY_AE}, we address the intra-particle vitrification process from a solid-state perspective. We consider a temperature $T < T_\textrm{g}$ where the colloids are amorphous solids with a finite shear modulus $G >0,$ and identify $T_\textrm{g}$ as the point at which a loss of mechanical stability (signaled by the vanishing of $G$) of the polymer chains results as a consequence of the reduction in the monomer connectivity driven by the Debye-Gr{\"u}neisen thermal expansion. By supplementing this picture with basic condensed matter physics consideration about thermal expansion, we investigate the repercussions of the intra-particle melting on the Hamaker constant $A_\textrm{H}$ of the suspension. We find $A_\textrm{H}$ to be inversely proportional to the thermal expansion coefficient $\alpha_T$ and, at the same time, directly proportional to $T_\textrm{g}.$ From the latter relation it follows that the intra-particle glass transition temperature $T_\textrm{g}$ can be used as a convenient parameter (typically more easily accessible in experiments than the Hamaker constant $A_\textrm{H}$) to estimate the degree of stability of a colloidal suspension with respect to particle coagulation. In particular, wthin DLVO theory, we find that the novel relationship of direct proportionality between $T_\textrm{g}$ and $A_\textrm{H},$ implies the CCIS to monotonically decrease as a function of $T_{\textrm{g}}.$ We conclude the Letter discussing how our theoretical predictions can be connected to the recent experimental findings of Scott et al. \cite{Priestley}.

A lot of research activity has been devoted to the general problem of the glass transition in the last decades \citep{DEB__STI,B&B}. While most available theories focus on the glass transition viewed ``from the liquid", i.e. study the dynamical arrest occurring by further cooling a (supercooled) liquid, for many applications it is more useful to follow the alternative approach of Zaccone and Terentjev \citep{GLASS_THEORY_AE}, which focuses on the glass transition viewed ``from the solid", i. e. studying the melting of an amorphous solid into a liquid. This perspective allows one to deal with the vitrification phenomenon by means of the tools of solid state science, in particular the nonaffine response theory of glasses \cite{Zaccone_book}, and to extract a (mechanical) signature of the glass transition from the temperature dependence of the  low-frequency shear modulus $G.$ More specifically, within nonaffine response theory, the shear modulus $G$ of an amorphous solid can be analytically linked to the average number of intermolecular contacts per particle (e. g. a monomer subunit group in a polymer chain) $z$ \citep{Scossa, Blundell}. In turn, $z$ can be connected to the absolute temperature $T$ and the glass transition temperature $T_\textrm{g}$ can be estimated by means of a generalized Born melting criterion \citep{Born_melting} as the critical value of $T$ at which $G$ vanishes, i. e. $G(T_{\textrm{g}})=0$ \citep{GLASS_THEORY_AE}.

In the presence of structural disorder, a solid lattice deforms under an applied strain very differently from well-ordered centrosymmetric crystals \citep{Alexander,lubensky, Tanguy, MacKintosh}. In the latter, indeed, the forces transmitted to every atom upon deformation by its bonded neighbors, cancel to zero. By contrast, due to the lack of local inversion symmetry, these forces do not balance in amorphous solids and can only be relaxed through a \textit{nonaffine} displacement which adds to the \textit{affine} one dictated by the macroscopic strain. In other words, as a consequence of the lattice disorder, atomic displacements in amorphous solids are strongly nonaffine. As nonaffine displacements are performed at the expense of the internal energy of the solid, the free energy of deformation of an amorphous solid under a shear strain $\gamma,$ has to be expressed as $F (\gamma) = F_\textrm{A} (\gamma) + F_\textrm{NA} (\gamma),$ where $F_\textrm{A} (\gamma)$ and $F_\textrm{NA} (\gamma)$ are the affine contribution (provided by the framework of Born-Huang lattice dynamics \citep{Born_Huang}), and the nonaffine contribution, respectively. Starting from $F (\gamma),$ an analytic expression for the shear modulus $G$ of an amorphous lattice can be derived by resorting to an eigenfunction decomposition of the nonaffine contribution. This has been done, for example, by Lema$\hat{\textrm{i}}$tre and Maloney \citep{Lemaitre1} and the result is given by $G = G_\textrm{A} - G_\textrm{NA} = G_\textrm{A} - \sum_{i} \uline{f}_i^{T} \sum_{j} \uuline{H}_{ij}^{-1} \underline{f}_j,$ where $\uuline{H}_{ij} = (\partial^2 \tilde{U} / \partial \uline{r}_i \partial \uline{r}_j)_{\gamma \to 0}$ represents the standard dynamical matrix of the solid \citep{AM}, $\tilde{U}$ the internal energy of the system, and $\uline{f}_i$ the force per unit strain acting on the atoms due to the shear deformation \citep{Lemaitre1}. As shown in Refs. \citep{Scossa,Blundell,GLASS_THEORY_AE}, assuming central-force interactions, $G$ can be evaluated analytically as
\begin{equation} \label{one}
\begin{aligned}
G &= G_\textrm{A}-G_\textrm{NA} = \frac{2}{5 \pi} \frac{\tilde{\kappa}}{R_0}  \phi (z-z_{c}),
\end{aligned}
\end{equation}
where $\phi$ is the packing fraction of the system, $R_0$ is the equilibrium lattice constant in the undeformed frame, and $\tilde{\kappa}$ is the lattice spring constant evaluated as $\tilde{\kappa} = \big( \partial^2 \tilde{U} / \partial \underline{u}^2_{ij} \big)_{R_0}.$ The nonaffine contribution is encoded in Eq. \eqref{one} in the term proportional to $z_c.$ More specifically, $z_c$  is a rigidity threshold defining the critical coordination at which the system is no longer rigid, because all the lattice potential energy is ``spent" on sustaining the nonaffine motions and no energy is left to support the elastic response to deformation.
In general, $z_c$ is proportional to the total number of degrees of freedom involved in the nonaffine energy relaxation, i. e. for purely central interactions in a $d$-dimensional space $z_c = 2d.$ In the $d=3$ case, it follows $z_c=6,$ consistently with purely central forces, $G \sim ( z- 6) $ \citep{Alexander,Scossa}.

It is worthwhile noticing that the direct contribution of thermal phonons to the elastic response is not included in the expression \eqref{one} of the shear modulus and will be neglected throughout this Letter, as for many materials it is very small compared with the contribution of nonaffinity \citep{GLASS_THEORY_AE}.


The crucial effect which controls the temperature dependence of the shear modulus $G$ of an amorphous solid is the change in the atomic connectivity $z$ due to the Debye-Gr{\"u}neisen thermal expansion.
Upon approaching the glass transition temperature $T_\textrm{g}$ from below, this effect is responsible for the loss of mechanical stability at $z = z_c.$ When heated, the volume $V$ of real molecular and atomic glasses increases. As a consequence, the atomic packing fraction $\phi$ decreases, an effect mediated by the thermal expansion coefficient defined as $\alpha_T \equiv \frac{1}{V} (\partial V / \partial T) = - \frac{1}{\phi} (\partial \phi / \partial T).$ Integrating this relation, $\phi$ can be seen to evolve with $T$ according to $\ln (1/\phi) = \alpha_T T +c,$ with $c$ an integration constant.
The average intermolecular connectivity of the monomers $z,$ instead, can be shown \citep{GLASS_THEORY_AE} to decrease while increasing $T$ according to the relation 
\begin{equation} \label{connectivity}
z(T) - z_c = \sqrt{\phi_\textrm{c} \big[ e^{\alpha_T (T_\textrm{g} - T)} - 1\big]},
\end{equation}
where $\phi_\textrm{c}$ is the (critical) packing fraction, reached by the system at the glass transition. Insertion of Eq. \eqref{connectivity} into Eq. \eqref{one}, reveals $T_{\textrm{g}}$ to correspond to the temperature at which the condition $z(T_{\textrm{g}})=z_c,$ causing the shear modulus $G$ to vanish, is reached. In other words, $T_\textrm{g}$ is the temperature at which the average number of total mechanical contacts on each atom $z$ becomes just sufficient to compensate non-affine relaxation and the glass consequently ceases to be an elastic solid \citep{Scossa,GLASS_THEORY_AE}.

The theory depicted above can be employed to deal with the vitrification within the nanoparticles dispersed in a colloidal suspension. As already stated, we consider the case of colloidal particles consisting of several polymer chains, each composed of $n$ monomer units with identical chemical structure. For this kind of systems, the framework of Zaccone and Terentjev allows one to find an estimate for the glass transition temperature $T_\textrm{g}$, by accounting for both covalent bonds along the chain and central-force London-van der Waals interactions. Indeed, when the already mentioned relation $\ln \big( 1/ \phi \big) = \alpha_T T + c$ is evaluated at the glass transition $(T_\textrm{g}, z_c,\phi_\textrm{c}),$ after linearization, one correctly obtains the Fox-Flory type \cite{Fox} relation between $T_g$, thermal expansion and molecular weight \cite{GLASS_THEORY_AE}:
\begin{equation} \label{two}
\alpha_T T_\textrm{g} \approx (1- c - \phi_c^{*} +2\Lambda) - \frac{2\Lambda}{n}.
\end{equation}
 When considering the values $n \approx 200,$ $\Lambda \approx 0.1,$ $\alpha_T = 2 \cdot 10^{-4} \textrm{K}^{-1},$ $\phi_c^{*}\approx 0.64$ as found in polymer glass \cite{Stachurski,Bonn,Hoy} and $T_\textrm{g} = 383 \ \textrm{K},$ a $G(T)$ profile in agreement with experimental data of Ref. \citep{Schmieder} follows from the insertion of Eq. \eqref{two} into the relation \eqref{connectivity}, and of the resulting expression, in turn, into Eq. \eqref{one} (see e. g. Fig. 4 of Ref. \citep{GLASS_THEORY_AE}).

In this Letter, we show that Eq. \eqref{two} can be used to estimate how $T_\textrm{g}$ affects the stability of a colloidal suspension. We recall that DLVO theory assumes the colloidal particles dispersed in a hosting solvent to interact through the pair potential \citep{Landau,V_O}
\begin{equation} \label{VDLVO}
V_\textrm{DLVO} (h) = V_\textrm{vdW} (h) + V_\textrm{R} (h),
\end{equation}
where the attractive vdW energy $V_\textrm{vdW} (h)$ and repulsive electrostatic interaction energy $V_\textrm{R} (h)$ are expressed as a function of the smallest surface separation distance $h.$ The latter is given by (see Fig. \ref{FIG1}$(c)$) $h = \tilde{r} - 2 a,$ where $a$ is the radius of the spherical colloids and $\tilde{r}$ their center-to-center separation. In particular, the vdW attraction acting between two spherical colloids can be written as \citep{Hamaker}
\begin{equation} \label{vdW_complete}
V_\textrm{vdW} (h) = \frac{- A_\textrm{H}}{12} \bigg( \frac{a^2}{h^2 + 2 a h} + \frac{a^2}{(a+h)^2} + 2 \log \frac{h^2 + 2 a h }{(a+h)^2} \bigg),
\end{equation}
from which it is clear that the Hamaker constant $A_\textrm{H}$ (being typically a positive quantity) measures how strong the attraction is among the dispersed colloids and hence contributes to quantifying the colloidal stability of the suspension. In the following, we prove that $A_\textrm{H}$ can be quantitatively connected to the thermal expansion coefficient $\alpha_T$ and, consequently, derive a link between intra-particle glass transition temperature $T_\textrm{g}$ and colloidal stability, exploiting Eq. \eqref{two}.

To establish a link between $\alpha_T$ and $A_\textrm{H},$ we focus on the interaction among monomers of different chains in each colloid and show how $\alpha_T$ can be computed once the non-covalent interaction pair potential $U (r)$ between two monomers separated by a distance $r,$ is known. To start, we notice that $\alpha_T$ can be written in terms of the \textit{linear} thermal expansion coefficient $\alpha_l,$ as $\alpha_T \approx 3 \alpha_l.$ In turn, $\alpha_l$ can be defined as
\begin{equation} \label{four}
    \alpha_{l} \equiv \frac{1}{\sigma}\frac{d \langle x \rangle}{dT}, 
\end{equation}
where $\sigma$ is the hard-core diameter of a single monomer (see Fig. \ref{FIG1}$(b)$), and $ \langle x \rangle$ is the average displacement of the monomers from their equilibrium positions, i.e. $\langle x \rangle$ represents the average value of the quantity $x \equiv r- r_\textrm{min}$ with $r_\textrm{min}$ being the bonding minimum of $U(r).$ Albeit belonging to a glassy out-of-equilibrium state, each monomer can be safely assumed to be locally at thermal equilibrium, such that $\langle x \rangle$ can be evaluated in the Boltzmann ensemble as
\begin{equation} \label{five}
    \langle x \rangle \equiv \frac{\int_{-\infty}^{\infty}x\,e^{-\beta U(x)} dx}{\int_{-\infty}^{\infty} e^{-\beta U(x)} dx},
\end{equation}
where $U$ is expressed as a function of $x,$ and $\beta \equiv (k_B T)^{-1}$ is the Boltzmann factor.

As shown by Kittel \citep{Kittel}, a convenient way to compute $ \langle x \rangle $ from Eq. \eqref{five} is to consider a  Taylor expansion of $U(x),$ which (up to fourth order) reads
\begin{equation} \label{six}
    U(x)=\zeta_2 x^{2}-\zeta_3 x^{3}-\zeta_4 x^{4}, 
\end{equation}
with $\zeta_2$, $\zeta_3$ and $\zeta_4$ real and positive coefficients. By assuming that the cubic and quartic anharmonic correction terms in Eq. \eqref{six} are small compared to $k_{B}T,$ it is possible to write $\int_{-\infty}^{\infty}x \ e^{-\beta U(x)} dx \approx  \int_{-\infty}^{\infty} x \ e^{\beta \zeta_2 x^2} ( 1+\beta \zeta_3 x^{3} +\beta \zeta_4 x^{4}) dx \approx (3 \pi^{1/2} \zeta_3 /4 \zeta_{2}^{5/2} ) \beta^{-3/2}$ and $\int_{-\infty}^{\infty} e^{-\beta U(x)} dx \approx \left( \pi / \beta \zeta_2 \right)^{1/2},$ respectively. As a consequence, Eq. \eqref{five} becomes
\begin{equation} \label{nine}
\langle x \rangle= \frac{3}{4} \frac{\zeta_3}{\zeta_2^{2}} k_{B}T.
\end{equation}
So, for a given interaction pair potential $U (r),$ once the coefficients $\zeta_2$ and $\zeta_3$ in the Taylor expansion \eqref{six} are known, the average displacement of the monomers $\langle x \rangle$ can be estimated through Eq. \eqref{nine} and the linear thermal expansion coefficient $\alpha_{l}$ can be in turn obtained by using Eq. \eqref{four}. Finally, the thermal expansion coefficient $\alpha_{T}$ can be obtained as $\alpha_{T} = 3 \alpha_{l}.$


In this Letter we assume that, within each colloidal particle, the non-covalent interaction pair potential $U (r)$ acting among monomers of different chains is dominated by attractive dispersion forces that decay as $r^{-6}.$ A (simple) expression, derived by London, for these attractive dispersion forces reads as \cite{London} 
\begin{equation} \label{london_potential}
U_\textrm{L} (r) \approx - \frac{3}{4} \frac{I \alpha_0^2}{r^6},
\end{equation}
where $I$ and $\alpha_0$ are the first ionization potential and the (material) polarizability, respectively. It follows that a good approximation for $U (r)$ is given by the Lennard-Jones potential $U_{\textrm{LJ}} (r)=4\epsilon [(\sigma / r )^{12}- (\sigma / r )^{6} ],$ where $\sigma$ is the hard-core diameter of each monomer (see Fig. \ref{FIG1}$(b)$) and the identification $\epsilon \approx 3 I \alpha_0^2 / (16 \sigma^6)$ is considered.



Once the non-covalent inter-monomer  interaction potential $U_\textrm{LJ} (r)$ is specified,  we follow the method introduced by Kittel and recalled above to compute the thermal expansion coefficient $\alpha_T.$ We consider a Taylor expansion of $U_\textrm{LJ}$ expressed as a function of $x \equiv r-r_{\textrm{min}},$ with $r_{\textrm{min}}=2^{1/6}\sigma$ the single (physically meaningful) minimum of $U_\textrm{LJ} (r).$ Up to third order, the expansion reads as
\begin{equation} \label{eleven}
\begin{aligned}
U_{\textrm{LJ}}(x) \approx & -\epsilon + \frac{36 \epsilon}{\sigma^{2}}2^{2/3}  x^{2} - \frac{756 \epsilon}{\sigma^{3}}\sqrt{2}  x^{3}.
\end{aligned}
\end{equation}
By comparing Eq. \eqref{eleven} to Eq. \eqref{six}, it follows that  $2^{3/2} \sigma^2 \zeta_{2}  = - 36 \epsilon$ and $\sigma^3 \zeta_{3} =  756 \sqrt{2} \epsilon,$ respectively, such that, from Eqs. \eqref{nine} and \eqref{four} (and recalling $\alpha_T = 3 \alpha_l$), one obtains
\begin{equation}  \label{twelve}
\begin{aligned}
 \alpha_T  = \frac{7}{ 2^{5/6}} \frac{k_B \sigma^6}{ I \alpha_0^2}.
\end{aligned}
\end{equation}
Expression \eqref{twelve} can be exploited to connect $\alpha_T$ to the Hamaker constant  $A_\textrm{H}.$ According to the London-Hamaker formula \citep{Hamaker}, $A_\textrm{H}$ can be written in terms of the parameters $I$ and $\alpha_0$ of the London potential \eqref{london_potential} as
\begin{equation} \label{thirtheen}
A_\textrm{H} = \frac{3 \pi^2}{4} I \alpha_0^2 \rho^2,
\end{equation}
where $\rho$ is the number of monomers per unit volume in the colloidal particles. From a comparison between Eq. \eqref{thirtheen} and Eq. \eqref{london_potential}, $A_\textrm{H}$ can be clarly seen to measure the strength of attraction within the colloids. By using the definition of packing fraction of spheres $\phi = \frac{4}{3} \pi ( \frac{\sigma}{2} )^3 \rho,$ insertion of Eq. \eqref{thirtheen} into Eq. \eqref{twelve} leads to
\begin{equation} \label{fourtheen}
\begin{aligned}
\alpha_{T} &= \frac{189}{2^{5/6}} \phi^2 \frac{k_B}{A_\textrm{H}},
\end{aligned}
\end{equation}
such that, from Eq. \eqref{two} it follows 
\begin{equation} \label{fifttheen}
\frac{A_\textrm{H}}{k_B T_\textrm{g}} = K,
\end{equation}
with 
\begin{equation}
K = \frac{189}{2^{5/6} } \phi_c^2 (1 - c -\phi_c),
\end{equation}
where we neglected the last term on the r.h.s. of Eq. \eqref{two} as it is small for long chains, and incorporated $2\Lambda$ into $c$ for ease of notation.

Equations \eqref{fourtheen} and \eqref{fifttheen} represent the main result of this Letter. They are molecular-level relationships, (to the best of our knowledge) never reported before, connecting $\alpha_T$ and $T_\textrm{g},$ respectively, to the Hamaker constant $A_\textrm{H}$ of a colloidal suspension. 
From Eq. \eqref{vdW_complete}, it is clear that $A_\textrm{H}$ controls the stability of a colloidal suspension as it quantifies the strength of the vdW attraction among the dispersed colloids. The effect of $T_\textrm{g}$ on colloidal stability follows from the direct proportionality relation between $T_\textrm{g}$ and $A_\textrm{H}$ expressed by Eq. \eqref{fifttheen}.
From a physical point of view, a larger value of the intra-particle glass transition temperature $T_\textrm{g}$ corresponds to a larger value of the Hamaker constant of the suspension and hence, from Eq. \eqref{thirtheen}, to a larger cohesive energy of London dispersion forces between the polymer chains inside each colloid. Indeed, the larger $T_\textrm{g},$ the larger the amount of energy required to ``break'' the cohesive non-covalent interactions, which is necessary to ``melt'' the polymer glassy chains within the colloid particle. In this sense, both $T_{g}$ and $A_H$ are measures of the strength of the non-covalent London-vdW forces between monomers inside the colloid, and they must be related to each other. This quantitative relation has been provided in this paper by Eqs.\eqref{fourtheen} and \eqref{fifttheen}.


\begin{figure}
\centering
\includegraphics[width = 1.0 \linewidth]{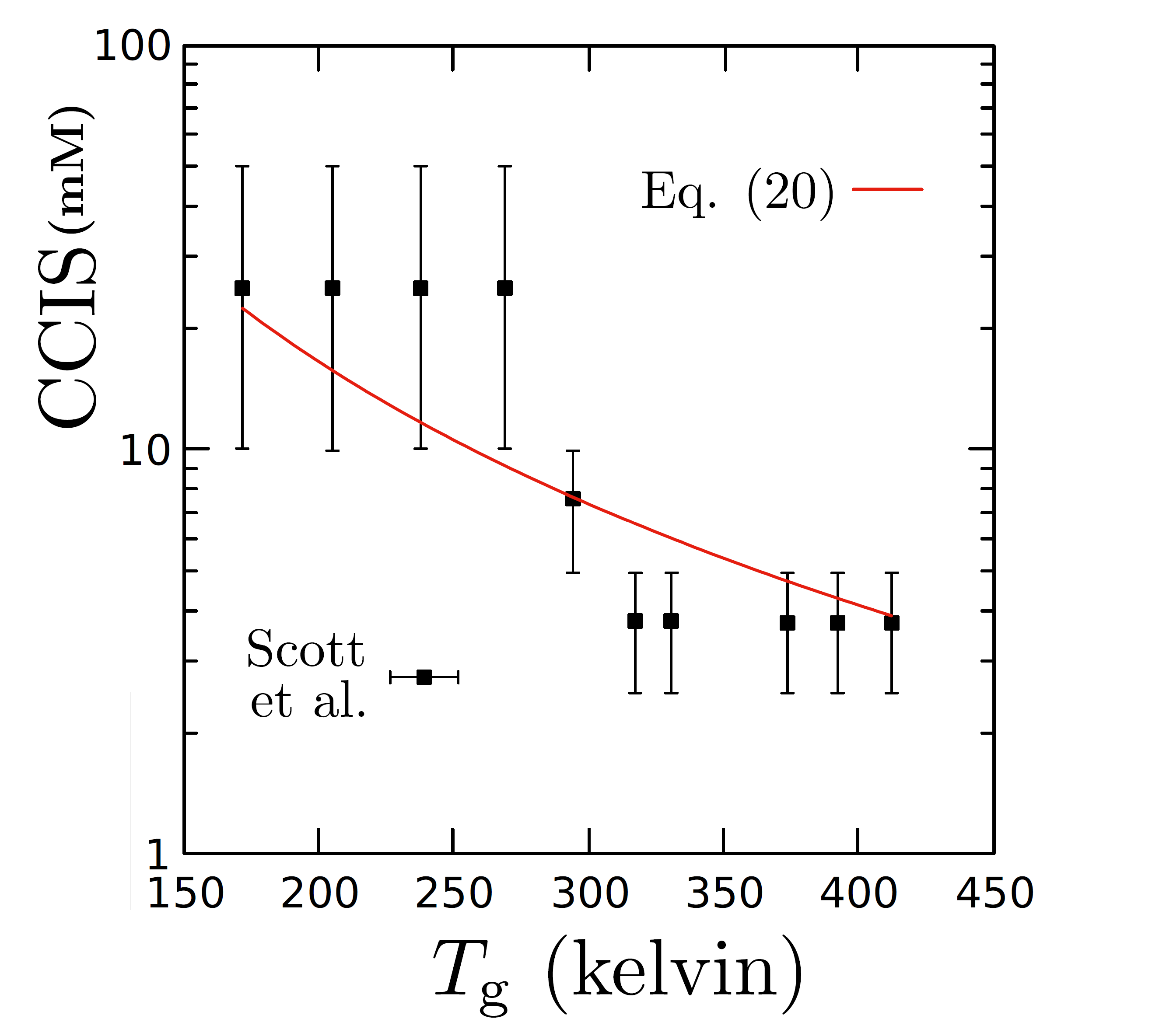}
	\caption{Critical coagulation ionic strength (CCIS) as a function of the inter-particle glass transition temperature $T_\textrm{g}.$ While points (and the corresponding error bars) are experimental data from Ref. \citep{Priestley}, full line represents Eq. \eqref{critical} with the proportionality constant $\eta$ obtained from a fit to experimental data.  Either our theory and experiments from Scott et al. predict the CCIS to be a decreasing function of $T_\textrm{g}.$ However, while theory predicts the decrease to be continuous, experiments predict the decrease to de discontinuous. Data from Ref. \citep{Priestley} are obtained when adding KCl salt for which the CCIS coincide with the critical coagulation concentration (CCC).}
		\label{FIG2}
\end{figure}

The critical packing fraction $\phi_\textrm{c}$ can be computed as \citep{GLASS_THEORY_AE, lappala} $\phi_\textrm{c} = \phi_\textrm{c}^* - \Lambda \cdot z_\textrm{co},$ where $\Lambda$ is a free parameter, $z_\textrm{co} = 2(1-1/n)$ is the average connectivity due to intra-chain covalent bonds and the packing fraction of non-covalently bonded particles $\phi_\textrm{c}^*$ can be reasonably assumed to coincide with the random close packing of a system of frictionless (hard) spheres, i. e. $\phi_\textrm{c}^* \approx 0.64$ \citep{Stachurski,Bonn,Bernal,zacconeRCP,anzivino}. With these values one obtains $A_\textrm{H} \approx  8.69 \times· 10^{-21} \ \textrm{J}$ in quite good agreement with experimental data on polystyrene particles \citep{experiments_HAMAKER}. Equation \eqref{fifttheen} represents also a new way to estimate the Hamaker constant once the rheological $G(T)$ of the glassy polymer, from which the colloid particles are made, is known.



Equation \eqref{fifttheen} can be finally used to estimate how $T_\textrm{g}$ affects the upper limit for salt addition beyond which particles begin to aggregate, i.e. how $T_\textrm{g}$ affects the CCIS of colloidal suspensions. We recall that, following Derjaguin and Landau \citep{Landau}, the CCIS can be computed directly from the interaction energy profile \eqref{VDLVO} provided that an expression for the repulsive term $V_\textrm{R} (h)$ is known. A simple expression for $\textrm{V}_{\textrm{R}} (h)$ can be derived by using the Debye-H\"{u}ckel approximation \citep{russel}, according to which one writes 
\begin{equation} \label{Dhukel}
\textrm{V}_{\textrm{R}} (h) \approx \textrm{V}_{\textrm{dl}} (h)= 2 \pi R \epsilon_0 \epsilon \psi_{\textrm{dl}}^2 e^{-\kappa h},
\end{equation}
where $\epsilon$ is the dielectric constant, $\epsilon_0$ is the vacuum permittivity, $\psi_{\textrm{dl}}$ is the diffuse-layer potential, and the inverse Debye length $\kappa$ is defined in terms of the ionic strength $I$ according to $\kappa^2 \equiv 2 q^2 I / (k_B T \epsilon_0 \epsilon).$ The CCIS corresponds to the value of ionic strength $I$ that satisfies the conditions 
\begin{equation} \label{conditions_minimize}
\textrm{V}_\textrm{DLVO} (h) =0 \ \  \ \textrm{and} \ \  \ \frac{d \textrm{V}_\textrm{DLVO} (h)}{ dh} =0,
\end{equation}
i. e. the CCIS represents the critical value of $I$ at which there is no energy barrier against aggregation \citep{Landau}. When the Debye-H\"{u}ckel approximation \eqref{Dhukel} is used for the repulsive potential $V_\textrm{R} (h)$ and the Derjaguin approximation $a \gg h$ \citep{Derjaguin_APPROX} is considered in the attractive vdW potential \eqref{vdW_complete}, the conditions \eqref{conditions_minimize} lead to \cite{russel,Borkovec}
\begin{equation}
\textrm{CCIS} = \frac{72 \pi}{e^{2}}\frac{1}{\lambda_{B}}\left(\frac{\epsilon \epsilon_{0} \psi_\textrm{dl}^{2}}{A_\textrm{H}}\right)^{2} \label{Landau},
\end{equation}
where $e$ is the Neper's number.
Upon inserting Eq. \eqref{fourtheen} into Eq. \eqref{Landau}, the CCIS can be seen to vary as a function of the intra-particle glass transition temperature, $T_\textrm{g},$ according to the relation
\begin{equation} \label{critical}
\textrm{CCIS} = \frac{72 \pi}{e^{2}}\frac{K^{-2}}{\lambda_{B}}\left(\frac{\epsilon \epsilon_{0} \psi_\textrm{dl}^{2}}{k_B T_\textrm{g}}\right)^{2}.
\end{equation}
Equation \eqref{critical} shows that the CCIS of a colloidal suspension decreases upon increasing $T_\textrm{g},$ with a law $\textrm{CCIS} \propto T_\textrm{g}^{-2}$. 
Equation \eqref{critical} allows us to test our theory against the recent experimental results of Scott et al. \cite{Priestley}. In particular, we consider data reported in Fig. 1 of Ref. \cite{Priestley} where the critical coagulation concentration (CCC), rather than the CCIS, of a suspension of electrostatic-stabilized polymer nanoparticles is measured by adding the hydrophilic KCl salt. In this case, the ionic strength $I$ and the concentration $c$ of the electrolytes present in the suspension are related by the condition \citep{Borkovec} $I=z^2 c,$ where $z$ is the valency of the electrolytes. Since the ions resulting from the addition of the KCl salt have $z=1,$ the CCC and the CCIS coincide in this case. 

A comparison between our theoretical prediction and the findings of Ref. \citep{Priestley} is presented in Fig. \ref{FIG2}, where experimental data are plotted as symbols while Eq. \eqref{critical} is represented by a full red line. The proportionality coefficient in Eq. \eqref{critical} is obtained by a fit to the experimental data. Either our theory and experiments from Scott et al. \cite{Priestley} predict the CCIS to be a decreasing function of $T_\textrm{g}.$ However, while theory predicts the decrease to be smooth, experiments rather display a sharper decrease, although this is difficult to ascertain due to the large error bars. Further investigation is required in the future to clarify this point and the possible influence of other effects that are not included in the above model. We observe that a different scaling can be obtained, namely $\textrm{CCIS} \propto T_{\textrm{g}}^{-2/3},$ if the charge-potential relationship $\sigma = \epsilon_0 \epsilon \kappa \psi_{\textrm{dl}}$ \cite{Borkovec} is used into Eq. \eqref{Dhukel}. However at the moment we cannot say which profile works better given the large error bars in the experiments.

To conclude, in this Letter we theoretically investigated how the stability of a colloidal suspension with respect to coagulation is influenced by the glass-transition temperature, $T_\textrm{g},$ of the suspended colloidal particles. We started by identifying $T_\textrm{g}$ with the point at which the glassy polymer chains within each colloid lose mechanical stability upon heating, as a consequence of the reduction of monomer connectivity driven by the Debye-Gr{\"u}neisen thermal expansion. We supplemented this picture with basic solid-state science considerations about thermal expansion, and established two novel relationships connecting the Hamaker constant $A_\textrm{H}$ of the suspension to the thermal expansion coefficient $\alpha_T$ and the intra-particle glass-transition temperature $T_\textrm{g},$ respectively. In particular, we found $A_\textrm{H}$ to be directly proportional to $T_\textrm{g}$ such that the latter quantity can be conveniently used as a key parameter (alternative to $A_\textrm{H}$) for controlling the stability of colloidal systems. The theory also provides an expression for the proportionality coefficient in terms of fundamental physical quantities.
Finally, within DLVO theory, we derived the critical coagulation ionic strength (CCIS) to be a monotonically decreasing function of the $T_{\textrm{g}}$ of the polymer.

The novel relations derived in this Letter may be useful for the design of colloidal materials whose stability can be tuned by varying the physical properties of the dispersed solid phase. For particles made of chemically complex materials, indeed, $T_{\textrm{g}}$ is commonly a much more accessible quantity compared to the Hamaker constant.

In future studies, it will be of significant interest to investigate the stability of of colloidal suspensions whose dispersed colloidal particles have an internal crystal structure, on the intra-particle melting temperature $T_\textrm{m}.$ Analogously to the intra-particle glass transition temperature $T_\textrm{g},$ $T_\textrm{m}$ can be connected to the thermal expansion coefficient $\alpha_T$ \citep{Granato}. We furthermore plan to extend the theoretical framework presented in this Letter to predict the effect on colloidal stability of $T_\textrm{g}$ in the presence of hydrophobic salts as those considered in Ref. \citep{Priestley} and to particles with non-spherical shape \cite{Tadmor_2001}.

\subsection*{Acknowledgments}
C.A. gratefully acknowledges financial support from Syngenta AG. A.Z. gratefully acknowledges funding from the European Union through Horizon Europe ERC Grant number: 101043968 ``Multimech'', from US Army Research Office through contract nr.   W911NF-22-2-0256, and from the Nieders{\"a}chsische Akademie der Wissenschaften zu G{\"o}ttingen in the frame of the Gauss Professorship program.

\bibliographystyle{apsrev4-1}
\bibliography{references}

\end{document}